# The Citation Field of Evolutionary Economics [*]

Wilfred Dolfsma[†] & Loet Leydesdorff[‡]

**Abstract:** Evolutionary economics has developed into an academic field of its own, institutionalized around, amongst others, the *Journal of Evolutionary Economics* (JEE). This paper analyzes the way and extent to which evolutionary economics has become an interdisciplinary journal, as its aim was: a journal that is indispensable in the exchange of expert knowledge on topics and using approaches that relate naturally with it. Analyzing citation data for the relevant academic field for the *Journal of Evolutionary Economics*, we use insights from scientometrics and social network analysis to find that, indeed, the JEE is a central player in this interdisciplinary field aiming mostly at understanding technological and regional dynamics. It does not, however, link firmly with the natural sciences (including biology) nor to management sciences, entrepreneurship, and organization studies. Another journal that could be perceived to have evolutionary acumen, *the Journal of Economic Issues*, does relate to heterodox economics journals and is relatively more involved in discussing issues of firm and industry organization. The JEE seems most keen to develop theoretical insights.

---

[*] We would like to thank three anonymous referee and the encouragement from editors Luigi Orsenigo and Uwe Cantner. The usual disclaimer holds nonetheless.
[†] University of Groningen, School of Economics and Business, PO Box 800, 9700 AV Groningen, The Netherlands, ph. (+31)50 363 2789. fax (+31)50 3637110, w.a.dolfsma@rug.nl
[‡] Amsterdam School of Communications Research (ASCoR), University of Amsterdam, Kloveniersburgwal 48, 1012 CX Amsterdam, The Netherlands, Tel.: +31-20-5256598, fax: +31-20-5253681 loet@leydesdorff.net; http://www.leydesdorff.net/

# 1. Introduction

A scientific community can, in line with Bourdieu (1992), be likened to a field where relations and behavior are institutionalized (Whitley 1984). One way in which academic activity is increasingly institutionalized is obviously in journals (Price 1985). Academic journals connect 'consumers' and 'producers' of scientific knowledge. Just exactly which knowledge is used by authors of articles in a particular journal, and where such knowledge is subsequently found relevant and is cited can sometimes be obvious to scholars but may also be surprising. This is true for the field of evolutionary economics and one of its main journal, the *Journal of Evolutionary Economics* (JEE) as well. In this contribution we provide a comprehensive analysis of the sources of knowledge used as input in the JEE as well as of the use that this knowledge is put to elsewhere.

Using methods developed in Social Network Analysis (SNA) and Scientometrics, it is thus possible to determine (i) which relations were shaped among journals in the field of evolutionary economics, and (ii) what knowledge is used more specifically in the JEE. The analysis provided may be particularly interesting if it presents a situation that is in some ways different from what could have been expected. We use aggregated citations among journals in this field in a number of different ways, instead of simply counting the number of cites or the Impact Factor (IF) of the *Journal of Evolutionary Economics*. We are interested in determining empirically which journals in the field of evolutionary economics occupy positions that turn them into indispensible links among academic fields and journals. Such an indispensable link established by a journal might but need not be reflected in its IF. In addition, we are interested in the extent to which the JEE can be considered as the *interdisciplinary* journal that it wanted to be when it was first published. Some links between journals can thus be expected, while others may not be obvious; some links are stronger than expected, while again other links between journals may be surprisingly absent. Links of a field with adjacent fields may, from a point of view of substantial discussion, be expected, but this does not mean that such links will materialize or have materialized in aggregates of references among journals.

Evolutionary Economics is briefly surveyed in the next section (Two). The goal is to characterize, rather than extensively or even exhaustively review (see Andersen 2009; Dopfer 2005; Hermann-Pillath 2009; Witt 2008), the academic field within economics that has taken explicit cues from biology and thermodynamics to develop ideas and conceptual frameworks that would explain economic phenomena. We aim to indicate how evolutionary economics, and particularly one of its main journal, relates to an outside discussion rather than primarily focus on how it has developed internally (Silva & Teixeira 2009). Section Three describes data used, and provides some descriptives with regard to the JEE as well. Some of the measures may not be familiar to economists as they are developed at the interface between SNA and science & technology studies. Discussing the proposed measure for interdisciplinarity, in particular, will feature in Section Four. Section Five discusses the findings before Section Six concludes the paper.

## 2. Characterizing the Evolutionary Economics field

Explicitly seeking inspiration from a number of different intellectual sources within economics itself, in the adjacent social sciences more generally, but also from the natural sciences (e.g., evolution theory, thermodynamics), evolutionary economists have claimed that their aim was to develop a quintessential interdisciplinary field of study (Hanusch 1991; Dosi 1991). Indeed, some authors have characterized evolutionary economics as a theoretical hybrid (Dopfer & Potts 2004). From within economics, cues from Austrian, Behavioral, Institutional, Post-Keynesian as well as Schumpeterian sub-domains are taken (Silva & Teixeira 2009).

Emphasizing concepts such as bounded rationality, heterogeneous agents, diverse learning paths, complex systems, disequilibrium and nonlinear dynamics, path dependence and lock-in (Dosi *et al.*, 2005; Lesourne 1991; Nelson 1995), this field, and particularly the JEE within it, positions itself firmly between the social sciences and reaches out to such natural sciences as thermodynamics, systems theory, complexity theory, cognitive science, computer science, and neuroscience (Boulding 1991; Dopfer & Potts 2004; Hanusch 1991). Stressing these themes and approaches, the JEE should be particularly situated to address questions on such topics as technical change, economic growth, industrial organization, learning dynamics (Silva & Teixeira 2009). The JEE is then expressly expected to "close the gap" thus identified (Hanusch 1991).

While a position between or at the interfaces of fields can make developing an "integrated approach" based on "clear principles" difficult to attain – as Silva & Teixeira (2009) observe about the field of evolutionary economics – it also may allow for new knowledge and insights to develop as well (Zitt 2005). Individuals that are able to tap into different kinds of sources of information will not only be in a strong position as knowledge broker (Burt 1992), but also in a position where these individuals can be expected to develop valuable new ideas themselves (Burt 2004). A similar argument will hold for the journals that institutionalize a scientific field where scholars from different backgrounds exchange knowledge.

Newly developing fields of scientific knowledge may lead to new journals or be accommodated within existing ones. For example, recent developments in the field of nanotechnology have evolved at interfaces among applied physics, chemistry, and the material sciences. New journals explicitly focusing on nanotechnology have emerged, but the developments in the field are also reported upon in incumbent journals such as *Nature* (Leydesdorff 2008b). What is more, the journals that focus on the newly developing field will actively refer to established journals, for instance to build on the knowledge that has developed there and to gain legitimacy. The delineation of a journal set is therefore by no means obvious. Existing classifications may have to be revised from the perspective of hindsight. While in substantive terms the knowledge produced by a journal may appear incohesive, when conceiving of the position of a journal and the field that it institutionalizes in relation to adjacent journals the picture can alter or become more fine-grained.

## 3. Citations and Impact

In a groundbreaking article, Garfield (1972) has proposed the aggregated number of citations for a journal as a measure of its importance in a field. Based on the number of citations to a journal, its Impact Factor can for instance be calculated.[1] An Impact Factor is considered by many as an indicator of a journal's importance. This proposal has, needless to say, provoked discussion (Bensman, 2007). Not all journals are included in the database that Thomson-Reuters / ISI produces. Two journals that could seem relevant for the field of evolutionary economics (cf. Silva & Teixeira 2009), for example, *History of Political Economy* and the *Review of Austrian Economics* are not included in the database. Admission is very selective, while criteria for inclusion into and exclusion from the database are not fully transparent (Garfield, 1990; Testa, 1997). When vying for (re-)inclusion, a steep hurdle needs to be taken by a journal, as it will need to be able to show it has an impact in the field. The quasi-IF calculated as part of the procedure underestimates the real IF as it does not include within-journal-self-citations that are included in the IF of journals that are included in the ISI database already.

---

[1] The Impact Factor (IF) for a journal in year $t$ is calculated based on a three-year period to $t$. The number of times articles published in $t-2$ and $t-1$ were cited in indexed journals during the year $t$ is divided by the number of "citable items." Citable items are articles, proceedings papers, reviews, and letters, but not editorials and obituaries.

As the need to evaluate scientific efforts and output grew (Gibbons et al. 1994), a number of additional measures have been proposed to characterize and evaluate journals. One is the total number of citations to a journal over the years. This measure of prestige disfavors newly established journals, however. Another is the *h*-index (Hirsch 2005), but it similarly favors accumulation over time. Most such measures use citation data provided by Thomson-Reuters / ISI. Most of the measures indicate a particular aspect of a journal (or an author), rather than important information about the journal as part of its field. Betweenness centrality, for example, can be used as an indicator of a journal's interdisciplinarity (Leydesdorff 2007b). We develop this below.

Citation patterns may change substantially from year to year. If a journal publishes more issues in a volume, the fluctuations tend to be smaller however. What is more, citation patterns differ greatly between fields. Most strikingly are the differences between the social sciences and the sciences. Among the social sciences there are important differences as well (Price 1972), sometimes following a business cycle of what topics are in demand. Comparing IFs across academic fields should be done with great care and comparison between journals can only be sensibly undertaken between journals in the same field (Garfield 1980). Even citation patterns between sub-fields within a single field may differ substantially, for example, among journals containing reviews, articles, or letters (Leydesdorff, 2008a). Some journals in a specific sub-field may also be more focused on the cutting edge of knowledge development in their field, or rather focus on reviewing the state of the art. Authors in other journals within the same sub-field may seek to relate to other fields (Goldstone & Leydesdorff, 2006). In some fields, for instance, scholars value monographs and edited volumes more than in others. If only because the information about such publications is not standardized, it is not included in the Thomson

Reuters/ISI database. Citations between journals in such domains can then give a very different indication of the development in a field than for one where journals are the major focal point for exchange of academic knowledge.[2]

ISI staff aggregates data among journals in the *Journal Citation Reports*. These reports contain three main indicators of journals: impact factors, immediacy indices, and subject categories. Subject categorization has remained the least objective among these indicators because the indicator is not citation-based but rather ISI-staff assigns journals to subjects on the basis of the journal's title, its citation patterns, etc.[3] Fields and sub-fields of sciences cannot (always) easily be determined, journals may change course over time, and journals published in different nations or with different publishing houses may not be easily classified. In addition, journals may in fact be in different fields at once (Boyack *et al.*, 2005): some journals indeed are included in the Social Science Citation Index (SSCI) and the Science Citation Index (SCI) databases. The problems of classification thus raised will be especially substantial, needless to say, for journals that aim to be between disciplines.

As the JEE is, or wants to be, in between the social sciences and the sciences, data from CD-Rom versions of the *Journal Citation Reports* of both the *Science Citation Index* as well as the *Social Sciences Citation Index* for the years 2000 to 2005 were collected. These are provided by Thomson-Reuters Scientific, formerly known as the Institute of Scientific Information (ISI). In particular we compare the situation in 2005 with that of 2004 in order to assess the enduring nature of the findings we present. Analyzing a combined SSCI and SCI database with citations between journals allows us to give a specific characterization of the current state of the field of evolutionary economics. We do so by taking the *Journal of Evolutionary Economics* (JEE) as a seed journal since citation environments only make sense locally (Leydesdorff 2006a; Dolfsma & Leydesdorff 2008). Our aim is to empirically determine to what extent the field of evolutionary economics has been able to meet this challenge of interdisciplinarity. For the year 2005 these two databases cover 6,088 and 1,747 journals, respectively. Since 301 journals are covered by both databases, a citation matrix can be constructed among (6,088 + 1,747 – 301) = 7,534 journals.

---

[2] Collecting such data about citations to (from) and between books oneself, to be added to the data about citations between journals, will, rather than complete the understanding of developments in a scientific domain, distort the understanding. Such data is not systematically collected while demarcations of fields is even less clear-cut for books than is the case for journals and cannot be adjusted expost based on empirical findings.

[3] In bibliometric research journals can be grouped either using the ISI subject categories (e.g., Leeuwen & Tijssen, 2000; Morillo *et al.*, 2003) or on the basis of clustering citation matrices (Doreian & Farraro, 1985; Leydesdorff, 1986; Tijssen *et al.*, 1987).

We compare the situation for the JEE with the other economic journal that is adamantly evolutionary in nature—the *Journal of Economic Issues* (JEI). Findings suggest that, in terms of citations between journals, the JEE in particular relates to the fields of geography and technology studies more than to (applied) game theory and finds itself in a more coherent field than the JEI. The JEI, however, relates to management journals—links that are remarkably absent from the picture for the JEE.

**4. Centrality**

To characterize a journal as interdisciplinary, one would need an indicator of "interdisciplinarity" for individual journals. To what extent are articles in a specific journal used in other academic domains, and to what extent do they draw upon different intellectual traditions themselves? A measure of interdisciplinarity at the level of journals could be a useful indicator of new developments one may expect in the near future (cf. Burt 2004). The social network literature suggests betweenness centrality as a suitable measure to indicate the extent to which a journal is positioned between academic fields operationalized as densely networked clusters among journals.

Freeman (1978/9) developed four broad concepts of centrality in a social network (Hanneman & Riddle, 2005; De Nooy *et al*., 2005). Some of these concepts of centrality themselves have variations (cf. Leenders *et al*. 2007). Some of these measures only make sense when a global perspective is taken, analyzing a full dataset, while others make sense also locally. Centrality can then be analysed in terms of

1. "degree": number of in- and outgoing information flows (ties, citations) from a node (local);
2. "closeness": the 'distance' of an agent from all other agents in a network (global);
3. "betweenness": the extent to which agent is positioned on the shortest path between any other pair of agents in the network (local); and
4. centrality in terms of the projection on the first "eigenvector" of the matrix (global).

*Degree* centrality is easiest to grasp as it is the number of relations a given node maintains. Degree can further be differentiated in terms of "in-degree" and "out-degree," or incoming and outgoing relations. In our case of a citation matrix, the aggregated references in the articles of a journal measure (weighted) out-degree centrality, and being cited can be considered as a measure of in-degree. As a measure of interdisciplinarity the number of citations or *degree* centrality is not the preferred indicator – journals might have many citations to it but all from the same academic field. Degree centrality is a measure that is highly sensitive to the absolute size of the focal journal, but it can be normalized as a percentage of the degrees in a network to control for scale effects.

*Closeness* centrality is defined as a proportion. First, the distance of a node from all other nodes in the network is counted. Normalization is achieved by dividing the number of other nodes by this sum total (De Nooy *et al*. 2005, p.127). Because of normalization, closeness centrality provides a global measure about the position of a vertex in the full network. Since academic practices, including citation patterns, can differ widely across fields, closeness centrality is not an attractive measure for interdisciplinarity.

Principal component and factor analysis decompose a matrix in terms of the latent *eigenvectors* which determine the positions of nodes in a network. *Eigenvector* centrality uses the factor loadings on the first eigenvector as a measure of centrality. While graph analysis begins with the vectors of observable *relations* among nodes (Burt, 1982), factor analysis positions nodes in terms of latent dimensions of the network. For example, core-periphery relations can be made visible using graph-analytical techniques, but not by using factor-analytical ones (Wagner & Leydesdorff, 2005).

*Betweenness* is a measure of how often a node is located on the shortest (geodesic) path between other nodes in the network, defined with reference to the local position of a vertex. Betweenness centrality thus measures the degree to which the node can control communication (Freeman 1978/9). Alternatively, the measure of betweenness centrality indicates what could happen to a network if a node (with a high level of betweenness centrality) were to be deleted from a network. In the absence of alternative, longer routes for knowledge to transfer (Leenders *et al*. 2007), one would expect such a network to fall apart into otherwise coherent clusters.[4]

---

[4] Betweenness centrality is normalized, by definition, as the proportion of all geodesics that include the vertex under study. If $g_{ij}$ is the number of geodesic paths between *i* and *j,* and $g_{ikj}$ is the number of these geodesics that pass through *k*, *k*'s betweenness centrality is:

One can expect that a journal which is "between" fields or groups of journals will show a different citation pattern from one that is in the middle of any particular academic field. Not necessarily belonging to the dense groups, but relating them, total citations to a journal that scores high for betweenness centrality may be low. Such a journal will load on different factors. Closeness is less dependent on relative positions between individual vertices because a vertex can be close to two (or more) densely connected clusters. Closeness can thus be expected to provide us with a global measure of *multi*-disciplinarity within a complete set, while betweenness provides us with a local measure of interdisciplinarity at specific interfaces (Leydesdorff 2007b).

In line with what is suggested in the relevant literature, we thus use *betweenness* centrality as it offers a measure for the extent to which a node in a network may control the information flow within a network (cf. Freeman 1978/9; Leydesdorff 2007b). While some have argued that information can flow more circuitously and not be channelled purposefully through the shortest route (Stephenson & Zelen 1989; Freeman *et al*. 1991), we believe that in the case of analyzing patterns of knowledge transfer at the level of academic journals the measure of betweenness centrality is to be preferred. Academics in this field will carefully determine which prior knowledge to build on (cite) when making their argument and when presenting their findings.[5] Furthermore, the more distant an academic journal is from another, the less easily reports that

$$\sum_i \sum_j \frac{g_{ijk}}{g_{ij}} \;,\; i \neq j \neq k$$

[5] MacRoberts & MacRoberts (1986. 1997, and 2010) cast doubt on this assumption based on research about citation patterns in specific fields of science (e.g., genetics). They find no substantial relation between a citation and actual influence, for instance. As in general the number of citations in each article has increased substantially, thus increasing the variety of sources drawn on and also increasing quality of research (Wuchty *et al.* 2007), these qualitative findings may no longer hold at the aggregated level, or at least not to the same degree. Also, given that the process of academic publication in the social science is slower than in the natural sciences, the findings of MacRoberts & MacRoberts may not be equally relevant for the field of evolutionary economics. There is more time to include acknowledgements of influence, and pressure to do so by editors and referees in the social sciences.

appear in it may be understood and the more 'translation' is required. We thus draw on social network analysis to analyse the position of nodes (i.e., journals) in a network in terms of betweenness centrality.

Centrality measures, contrary to impact factors, are sensitive to both the size of a journal as well as of a field. Correlations between different centrality measures can then be spurious: a large journals (e.g., *Nature*) which one would expect to be "multidisciplinary" rather than "interdisciplinary," might generate a high betweenness centrality simply because of its high degree centrality (number of citations to it). Normalization of the matrix for the size of patterns of citation can suppress this effect. There is increasing consensus in the information sciences that normalization in terms of the cosine and using the vector-space model provides the best option in the case of sparse matrices (Ahlgren *et al*. 2003; Salton & McGill 1983; Leydesdorff & Vaughan 2007). Salton's cosine is defined as the cosine of the angle enclosed between two vectors *x* and *y* (Salton & McGill, 1983):

$$\cos(x, y) = \frac{\sum_{i=1}^{n} x_i y_i}{\sqrt{\sum_{i=1}^{n} x_i^2} \sqrt{\sum_{i=1}^{n} y_i^2}} = \frac{\sum_{i=1}^{n} x_i y_i}{\sqrt{(\sum_{i=1}^{n} x_i^2) * (\sum_{i=1}^{n} y_i^2)}}$$

Using the cosine for the visualization, a threshold has to be set because the cosine between citation patterns of locally related journals will almost never be equal to zero. The cosine is very similar to the Pearson correlation coefficient, except that the latter measure normalizes the values of the variables with reference to the arithmetic mean (Jones & Furnas, 1987). The cosine normalizes with reference to the geometrical mean. Unlike the Pearson correlation coefficient, the cosine is non-metric and does not presume normality of the distribution (Ahlgren *et al*., 2003).

Given that citation behaviour tends to be highly context specific, it is of the utmost importance that the local environment of a particular journal—the seed journal—is analysed. Extraction of a relevant set of journals in a journal's neighbourhood is given by including all journals citing or cited by the specific journal in one's initial analysis. Subsequently imposing a threshold level below which connections to or from a seed journal are not presented in a figure enhances a visualization. Choice of a seed journal is formal and does not affect the measures calculated in this study.

These measures, their further statistical elaboration, and visualization of the accompanying networks are conveniently combined in the software packages like UCINet (Bonacich, 1987; Borgatti *et al*., 2002) and Pajek (De Nooy *et al*. 2005). We have used the data analysis software package Pajek for the visualizations and analyses. The figures presented include citation relations among journals that contribute more than 1% to the total citations of a seed journal. This threshold is used in order to produce readable representations. The *y*-axis of a node / circle indicates the logarithm of the number of cites the journal receives in this environment, while the *x*-axis corrects for self-citations. Thus, the larger the node for a journal, the more citations it receives in this citation environment. The rounder a node, the fewer self-citations it has. Highly elliptical nodes thus can be considered as journals that may have momentum, focusing on internal knowledge development. Alternatively, it may be that such journals have relatively little to offer to scholars publishing in other journals.

## 5. Findings

Since its start in 1991, the JEE has witnessed an astounding growth. Despite its relative youth, economists in increasing numbers have come to recognize the journal as a legitimate and important place to have one's work published and to become aware of important developments in the field. This is reflected in its impact factor, which has averaged 0.6 over the 2003-5 period, and has risen to 1.26 for the year 2008. This IF is substantial for the social sciences, but less so for a science journal. The JEE has thus waxed, and seemed to wane at times: perhaps our analysis sheds light on this development.

Table 1 provides descriptive statistics of normalized citation patterns (cosine>0.2) for journals in the JEE environment. The citing data is not used in a figure that is included in this paper, but the cited data is (Figure 1). Some indication of the relative weight of citations in this field can be gained (the *y* axis), as well as a quantitative indication of the extent to which articles in journals cite other work that has appeared in the same journal (self-citations, or the *x* axis). In terms of number of citations counted in this field, the JEE is a not among the larger players. The JEE does not self-cite frequently when compared to some other journals. The JEE thus seems to be a relatively open journal that is not overly focused on developing knowledge within its own group of scholars only.

**Table 1: Journals having a citing or cited relation to JEE, 2005**

|  | Citing | | Cited | |
|---|---|---|---|---|
|  | *x* | *y* | *x* | *y* |
| *Econometrica* | 3.14 | 10.14 |  |  |
| *AmEconRev* | 12.84 | 22.66 |  |  |
| *EconJ* | 8.49 | 10.81 | 7.14 | 8.53 |
| *IndCorpChange* | 5.57 | 7.74 | 3.05 | 4.36 |
| *JEconBehavOrgan* | 10.47 | 12.96 | 2.43 | 3.93 |
| *JEconPerspect* | 4.75 | 6.19 |  |  |
| **JEvolEcon** | **3.71** | **4.45** | **1.91** | **2.36** |
| *JPolitEcon* | 4.16 | 7.08 |  |  |
| *ResPolicy* | 7.27 | 19.96 | 7.19 | 13.59 |
| *EnvironPlannA* |  |  | 2.06 | 4.99 |
| *RegStud* |  |  | 4.88 | 8.31 |
| *TijdschrEconSocGe* |  |  | 0.46 | 0.90 |

| Journal | | | x | y |
|---|---|---|---|---|
| *EurPlanStud* | | | 1.13 | 2.24 |
| *EurUrbanRegStud* | | | 0.85 | 1.29 |
| *WorldDev* | | | 2.28 | 7.33 |
| *Scientometrics* | | | 0.74 | 8.44 |
| *TechnolForecastSoc* | | | 0.59 | 3.10 |
| *InfEconPolicy* | | | 0.06 | 0.33 |
| *JEconDynControl* | | | 1.19 | 2.96 |
| *JeconTheory* | | | 3.30 | 6.36 |
| *Jecon* | | | 0.15 | 0.34 |
| *SmallBusEcon* | | | 1.07 | 3.60 |
| *OrganStud* | | | 0.46 | 3.47 |
| *EcolEcon* | | | 0.50 | 5.97 |
| *CambridgeJEcon* | | | 2.18 | 3.45 |
| *JeconIssues* | | | 0.50 | 3.25 |
| *JahrbNatlStat* | | | 0.00 | 0.36 |
| *PolitEkon* | | | 0.00 | 0.53 |

Citations between journals. *y* represents gross citations (citing, and cited), while *x* is net of self-citations. Source: Thomson Scientific Journal Citation Reports, 2005; calculations by authors. Notes: Only journals that contributed over 1% of total cites to (or from) JEE are included in table.

Looking at Table 1, it is striking how little overlap there is for the JEE between the journals it cites and the ones it receives citations from. The journals that the JEE cites tend to be mainstream technical journals in the area of economics, true to what Dosi (1991) anticipated. These are journals that are well-established and have a high IF. If the JEE itself is cited by mainstream journal, these have a base in or relatively strong link to Europe, as evidenced from the membership of the editorial boards. Such journals are the *Economic Journal* and the *Journal of Economics*. The JEE itself cites American based mainstream journals such as the *American Economic Review*, the *Journal of Political Economy*, *Econometrica*, but that favor is not returned yet.[6]

---

[6] This is in line with what Merton (1968) suggests about social processes in the development of science.

The JEE indeed is an interdisciplinary journal to a large extent. As can be expected, it relates to the fields of technology studies. Of the two other journals conceptually close to evolutionary economics (Silva & Teixeira 2009), the link to *Industrial and Corporate Change* (ICC) is stronger, however, than that with *Research Policy*. JEE's main advantage in comparison to one of the behemoths of the field, *Research Policy* (ResPol), is its link to economics and economic theory. While such links at the same time may stand in the way of developing the kind of interdisciplinary posititon envisioned at its start, they currently provide the JEE a strong position. The JEE also is cited often by journals in the field of regional studies. The JEE may thus provide the necessary translations between the knowledge produced in mainstream economics journals and journals in the applied domains of regional studies or technology studies. ResPol relates strongly to regional science journals as well and, given its size, could well seek stronger direct lines with economics. Such a development could weaken JEE's position.

The journals that the JEE cites itself tend to be relatively open journals in the sense that they have relatively few self-citations. The journals that the JEE receives cites from are smaller journals in terms of IF and degree centrality. Journals in which papers published in the JEE are cited have quite a few self-citations: they may be focused on internal knowledge development to a larger degree.

**Figure 1:** *Journal of Evolutionary Economics*, cited, 2005

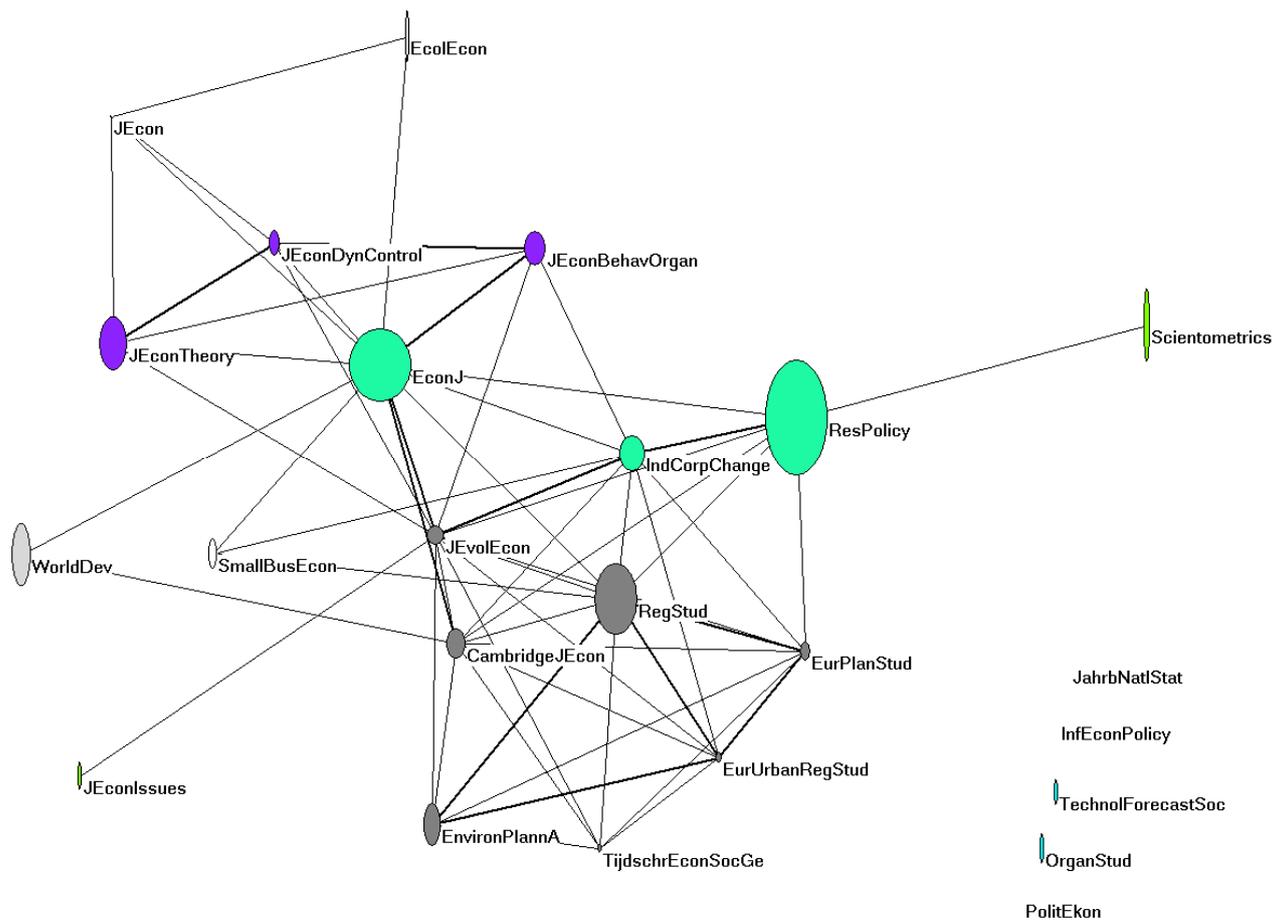

Source: Thomson ISI Journal Citation Reports, 2005; own calculations using Pajek software.
Notes: *Journal of Evolutionary Economics* as seed journal. Citations between journals are normalized (cosine>0.2) to enhance visualization. Size of node represents (the logarithm of) the number of citations; *x*-axis corrects for self-citations. Thickness of lines/edges represents the strength of the association.

Of the six main groups that can be distinguished conceptually in evolutionary economics (Silva & Teixeira 2009) – (1) institutionalist in the tradition of Veblen and Commons, (2) Schumpeterian, (3) Austrian, (4) evolutionary game theorists, (5) Santa Fe, and (6) "various writers such as Smith, Marx and Marshall" – the actual links to the second and the fourth appear to be best developed. Unfortunately, Austrian economics journals do not feature in the ISI list of journals, so taking a network perspective does not allow one to indicate what relation, if any, exists between the JEE and Austrian economics. The JEE appears to focus mainly on 'technical', game theory contributions as well as applications in the spheres of industry and technology dynamics. A journal that is argued to be central to JEE – *Games and Economic Behavior* (Silva & Teixeira 2009) – is not found in the pictures for 2005 or 2004 however. This journal seems to be more closely aligned with mainstream journals (see Appendix A). Another prediction (Hanusch 1991) has not come true so far either: the JEE is not immediately relating to physics or biology, at least not in terms of substantial number of citation relations to journals in these fields. Within the social sciences, the JEE relates to geography, but not to other social sciences such as (cognitive) psychology or sociology. ResPol has indeed also to a large degree severed its links with the more sociological approaches to the study of science and technology it once shared with the journal *Social Studies of Science*.

Table 2 presents different centrality measures for journals relevant to the JEE, sorted (primarily) by Betweenness Centrality (cited). The betweenness centrality measure is, as we argued above, important to understand the extent to which a journal is interdisciplinary. The measures have been calculated as explained in section 4. Those in columns 3 through 5 on the basis of the matrix of citations among all 7,525 journals included in the (*Social*) *Science Citation Indices*, and those in columns 1 and 2 for the journals that relate to the JEE as a seed journal. Obviously, the former set of journals is larger than the ones presented; we have only presented those that were also included in the latter group of journals. It shows how Impact Factor (in SNA parlance related to the measure of degree centrality) can provide a different impression than betweenness centrality: a journal such as the *Journal of Economic and Social Geography* (TESG) receives quite a few cites but is not strongly positioned between different academic fields relevant for knowledge development in the domain of evolutionary economics as indicated by its low betweenness centrality measure.

The high betweenness centrality score indicates that the JEE is an indespensible journal in the sense that it relates fields and discussions that might otherwise remain unrelated. The JEE itself is ranked surprisingly higher than one might expect on the basis of its Impact Factor – in the top 5 it is the journal with the lowest IF. This finding is not a mere artifact of the method chosen of taking the JEE as a seed journal as similar analyses for other journals that the authors could provide to any reader interested will attest. The JEE is not the highest ranked in terms of betweenness centrality in its field. In addition, the gap with number three is quite substantial. The high betweenness centrality might explain the increase in IF that the JEE has experienced in more recent years. JEE has not suffered so far from the recent appearance on the academic scene of *Industrial and Corporate Change* (ICC) in particular. ICC, while it has a commendably high Impact Factor, is not as indispensable a journal for the purpose of knowledge exchange between separate but not necessarily related knowledge domains as the JEE is.[7]

**Table 2: Centrality Measures of Selected Journals in the JEE field included in the *Science Citation Index* and *Social Science Citation Index*, 2005**

| Journal | Betweenness Centrality (%) (Local) (1) | Degree Centrality (#) (2) | Degree Centrality (in) (3) | Degree Centrality (out) (Global) (4) | Impact Factor (5) |
|---|---|---|---|---|---|
| *EconJ* | 16.90 | 24 | 316 | 118 | 1.44 |
| ***JEvolEcon*** | **15.87** | **26** | **41** | **48** | **0.53** |
| *ResPolicy* | 6.93 | 14 | 127 | 154 | 1.84 |
| *CambridgeJEcon* | 4.64 | 20 | 74 | 109 | 0.77 |
| *RegStud* | 3.72 | 20 | 93 | 151 | 1.53 |
| *IndCorpChange* | 2.81 | 18 | 78 | 91 | 1.08 |
| *EurPlanStud* | 0.71 | 16 | 27 | 84 | 0.51[a] |
| *JEconDynControl* | 0.68 | 10 | 102 | 70 | 0.69 |

---
[7] This was confirmed in an analysis using *Industrial and Corporate Change* (ICC) as a seed journal.

| | | | | | |
|---|---|---|---|---|---|
| *JEconTheory* | 0.68 | 10 | 165 | 44 | 0.91 |
| *JEcon* | 0.40 | 8 | 26 | 58 | |
| *JEconBehavOrgan* | 0.26 | 10 | 156 | 132 | 0.78 |
| *EurUrbanRegStud* | 0.16 | 7 | 29 | 80 | 0.75 |
| *TijdschrEconSocGe* | 0.00 | 12 | 30 | 83 | 0.61[a] |
| *EnvironPlannA* | 0.00 | 12 | 136 | 222 | 1.37 |
| *SmallBusEcon* | 0.00 | 6 | 43 | 116 | 0.53 |
| *WorldDev* | 0.00 | 4 | 228 | 165 | 1.50 |
| *JEconIssues* | 0.00 | 2 | 30 | 81 | 0.31 |
| *EcolEcon* | 0.00 | 4 | 175 | 249 | 1.18 |
| *Scientometrics* | 0.00 | 2 | 71 | 123 | |
| *OrganStud* | 0.00 | 0 | 98 | 130 | 1.28 |
| *TechnolForecastSoc* | 0.00 | 0 | 37 | 58 | 0.81 |
| *InfEconPolicy* | 0.00 | 0 | 8 | 37 | |
| *JahrbNatlStat* | 0.00 | 0 | 3 | 52 | |
| *PolitEkon* | 0.00 | 0 | 3 | 42 | |

Source: Thomson Scientific Journal Citation Reports; local centrality measures – own calculation.
Notes: Journals have been sorted by Betweenness Centrality (local), then Degree centrality (local), then Degree centrality (out; global); [a] 2006.

Given the Schumpeterian legacy that the JEE has, surprisingly tenuous links exist with entrepreneurship journals, or management journals more generally. As the field of entrepreneurship is currently experiencing a boom, this is where the JEE could expect a positive effect beyond the connection to *Small Business Economics* in the near future. Indeed, the JEE could well be thought to provide the indispensible link between management and entrepreneurship journals on the one hand, and technology development journals on the other hand. This is where it could then also challenge its main competitor in this respect: the *Journal of Economic Behavior and Organization*.

The JEE links to less mainstream, or even heterodox journals such as the *Journal of Economic Issues* (JEI) or the *Cambridge Journal of Economics* (CJE) are more tenuous, despite the expectation from the first editor in his editorial that there would be such links (Hanusch 1991). This may be due to the fact that the mainstream itself is changing (Davis 2006). Nevertheless, only CJE and JEI enter Figure 1. In 2004 there was not even a direct link between the JEE and the JEI that passed the threshold value for inclusion in a figure (see Appendix B). It would appear from this that these two journals represent different perspectives on economic evolution.

This impression of separate perspectives on economic evolution is partly confirmed in Figure 2. To some degree the same other journals are present in Figure 2 as in Figure 1 – CJE and *Research Policy* come to mind. In this analysis the JEE, rather than seed journal JEI, has the highest betweenness centrality score (together with CJE). For the *Journal of Economic Issues* the academic field is less tightly connected than for the JEE, and within it the JEI's position is less central than that for the JEE. Some connections do not survive the threshold imposed to make sure a visualization remains tractable. The JEI is cited by a cluster of journals that focuses on organizations, however, a field that the JEE could be expected to play a role in but does not relate to. Yet, this connection is tenuous. The JEI, sponsored by the Association for Evolutionary Economics (AFEE), seems to have a view of evolutionary change in the economy that differs from the one drawing more directly from Charles Darwin mostly endorsed in the JEE (*Journal of Evolutionary Economics* 2006). In the JEI agency and purpose play a more prominent role (if sometimes implicitly acknowledged) and the unit of analysis is that of institutions (cf. Bush 1987; Dolfsma 2009; Hodgson 2004). These may be reasons for the relatively close connection with general management journals, a field where institutional theory has come to be increasingly referred to (see, e.g., Scott 2001).

**Figure 2:** *Journal of Economic Issues*, cited, 2005.

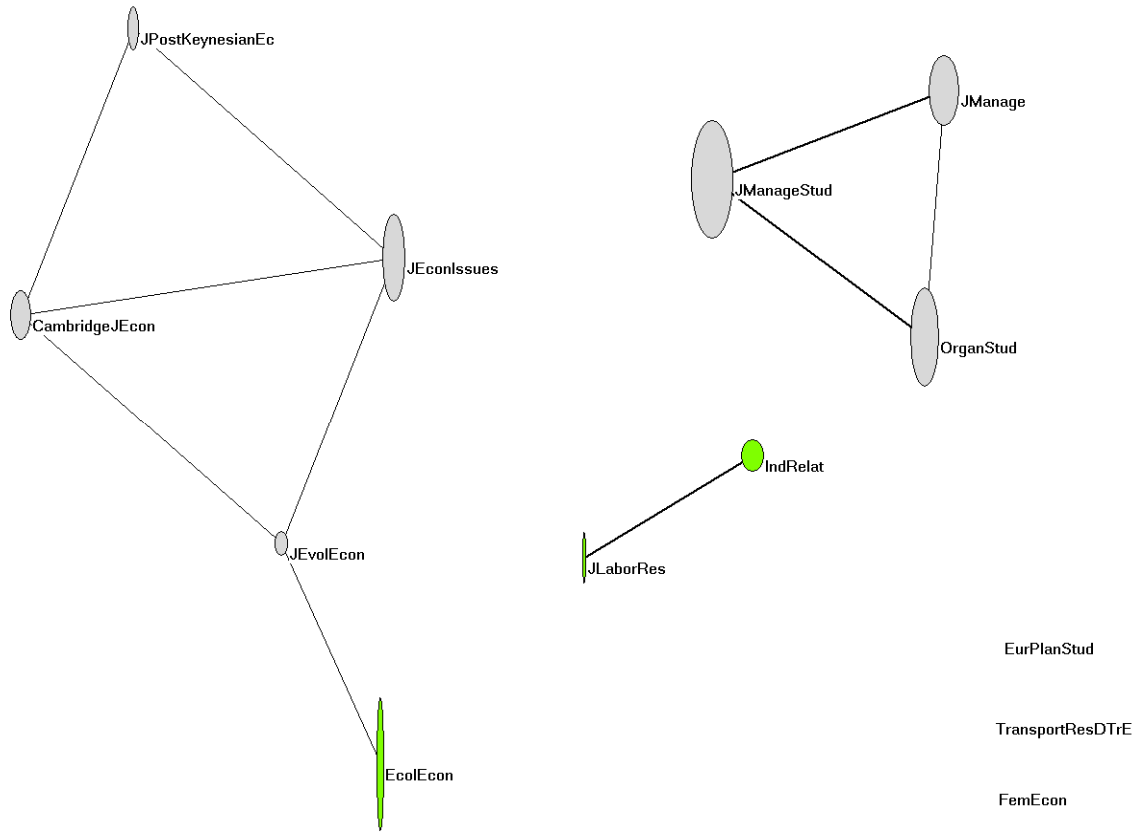

Source: Thomson ISI Journal Citation Reports, 2005; own calculations using Pajek software.
Notes: *Journal of Economic Issues* as seed journal. Citations between journals are normalized (cosine>0.2) to enhance visualization. Size of node represents number of citations (normalized); *x*-axis corrects for self-citations. Thickness of lined/edges represents the strength of the association.

## 6. By Way of Conclusion

We would like to end this contribution in a somewhat unusual way, that some may possibly perceive as supercilious, by offering some suggestions to JEE editors, authors, and readers. These suggestions obviously are only tenuously related to the substantial arguments of evolutionary economics as they are based on the citation analysis for the JEE as presented in this paper. Furthermore, our analysis is at the level of aggregated citation relations among journals only for reasons spelled our earlier. Citations are included by authors at the micro-level in articles published in the many journals in the (Social) Science Citation Index, and thus this measure is somewhat removed from the intentional behavior of these agents. Nevertheless, communities of agents play an active part showing behavioral patterns that gradually affect the citations between journals and thus the network configuration empirically analyzed in this paper.[8]

First of all, the JEE might want to seek stronger citation links with journals in the neighborhood that are obviously influential in having a substantial node. Since there is surprisingly little overlap between what others journals the JEE cites and where it receives its own citations from, there are opportunities to help contributing to a stronger local cluster in which the JEE could take a central position. One way of bringing this about would be to link closer with *Research Policy* and *ICC*, which would involve redirecting the current thrust of contributions slightly away from 'high [game] theory' towards publishing empirical research (cf. Silva & Teixeira 2009).

Even when their autonomy is circumscribed at least in the analysis we offer here, actors involved in the JEE, and most importantly its editors, might seek links with journals that do not disproportionately cite within their own journal. Such links could (substantially) limit its network horizon beyond these journals (Van Liere 2007). It would appear that the JEE has indeed managed to position itself in the relevant network as a structural hole, an indispensable go-between, or translator of ideas, between otherwise separate fields. Yet, this can be further exploited by emphasizing other elements of its Schumpeterian roots that the JEE has, for instance by indicating that it is well-positioned to deal with uncertainty, indeterminacy and disequilibrium, as anticipated by Hanusch (1991), Dosi (1991) and Boulding (1991) in the

---

[8] Our observations, we suggest, thus, go beyond what actions some propose based on an understanding of the technicalities of calculating Impact Factors. Such recommendations include: limiting the number of relatively short articles as that deludes the denominator, publishing the articles one expects to draw many cites in the first issue(s) of the year as they can then collect citations for a longer period, encouraging citation of JEE papers that have been (recently) published for newly submitted papers, or commissioning review and survey articles as they will feature in many an introduction for subsequent papers (e.g. Cookson & Cross 2006, 2007).

Journal's first issue. It could then further strengthen its position vis-à-vis such areas as technology studies and regional studies as well as studies of management and entrepreneurship.

Given the increasing extent to which economic domains other than technology are also characterized by uncertainty, indeterminacy, and disequilibrium (e.g., financial markets), the JEE could seek links to novel domains. As the JEE seeks to relate to fields that are not obviously specified by technology, it may need to reconsider its position in regards to evolutionary processes, however, focusing less on irreversibilities and more on anticipation (Andersen 1994, Leydesdorff 2006), recognizing that there can be break-outs from a lock-in (Dolfsma & Leydesdorff 2009). The JEE could also be an indispensible link between technology studies and the management literature. This should not involve an imposing change of focus as close affinity to the Schumpeterian legacy can be maintained, most obviously when emphasizing the theme of, for example, entrepreneurship.

Appendices

A – Citation Environment for *Games and Economic Behavior*, cited 2005

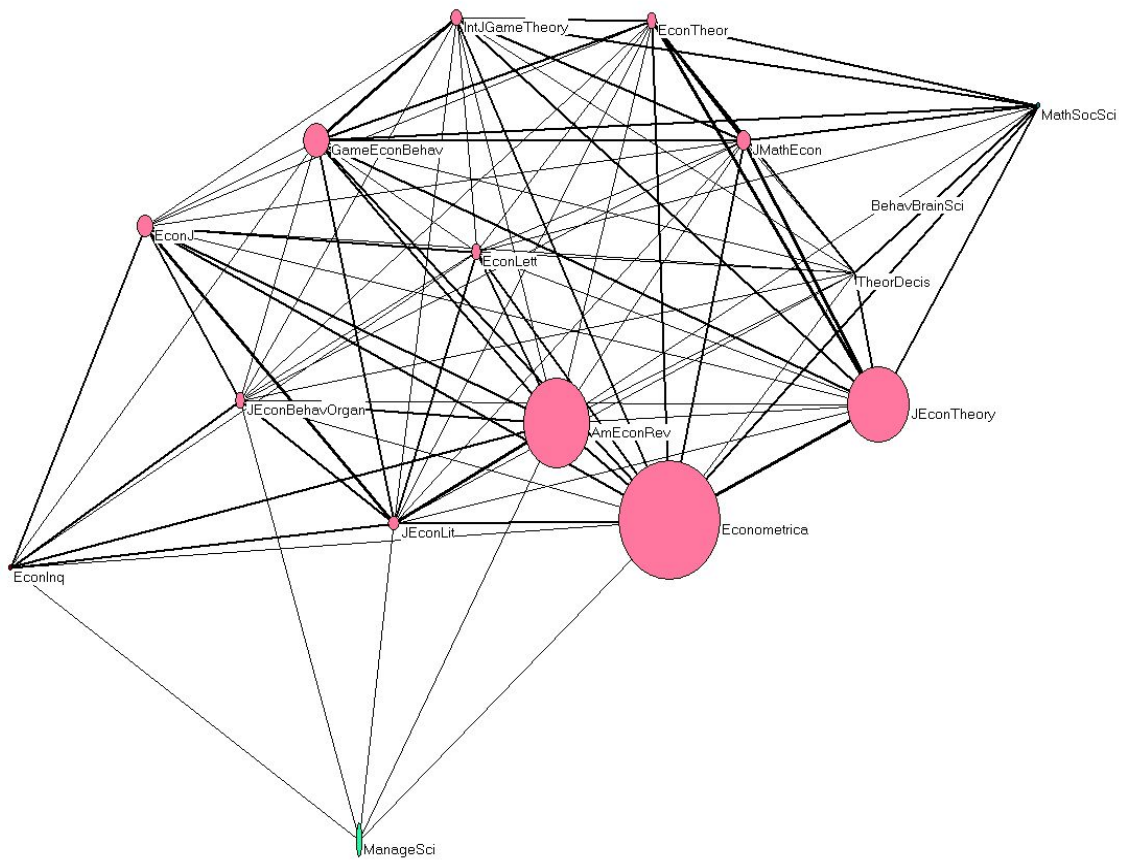

Source: Thomson ISI Journal Citation Reports, 2005; own calculations using Pajek software.
Notes: *Games and Economic Behavior* as seed journal. Citations between journals are normalized (cosine>0.2) to enhance visualization. Size of node represents number of citations (normalized); *x*-axis corrects for self-citations. Thickness of lined/edges represents tie strength.

B: *Journal of Evolutionary Economics*, cited, 2004

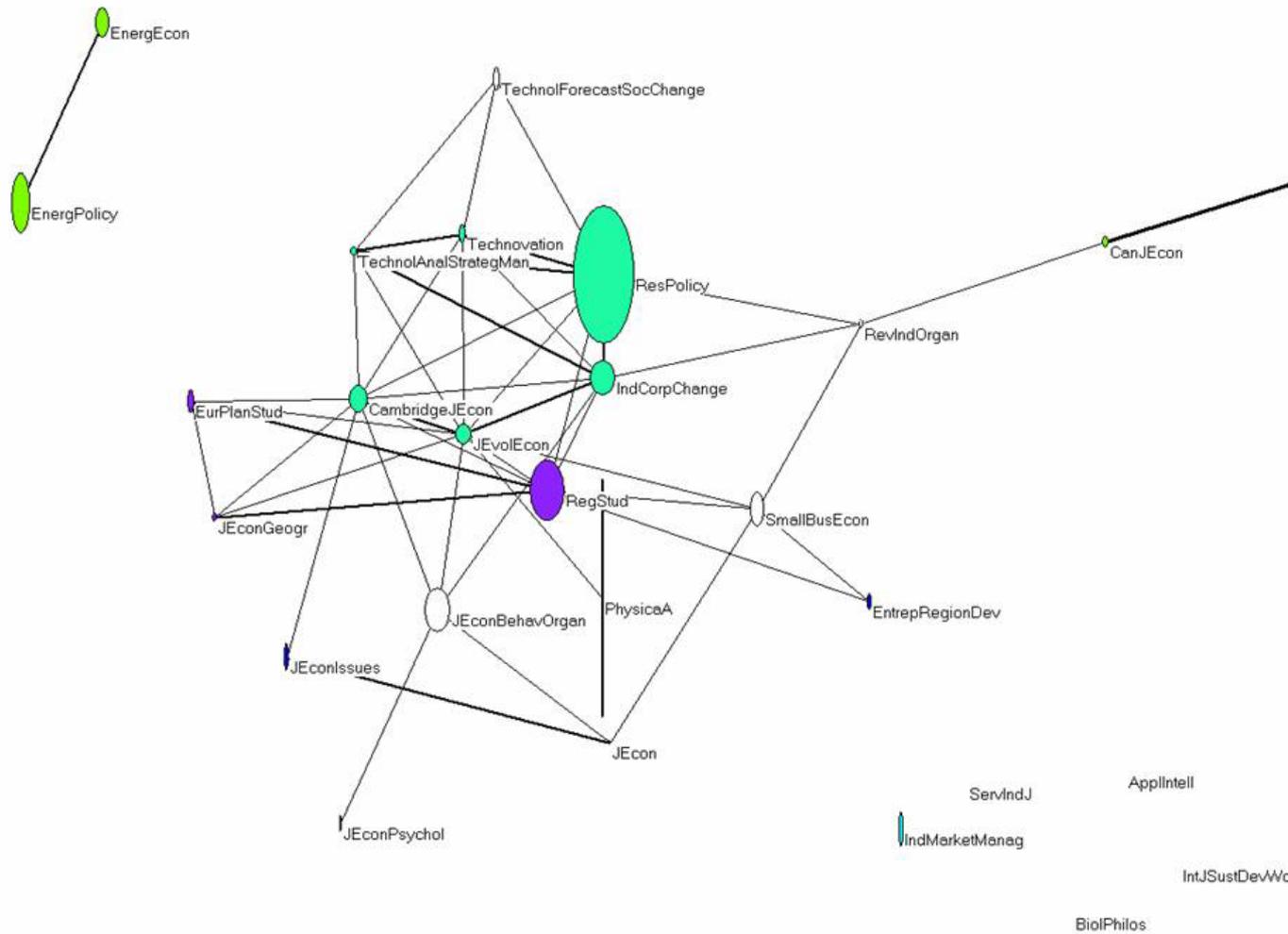

Source: Thomson ISI Journal Citation Reports, 2004; own calculations using Pajek software.
Notes: *Journal of Evolutionary Economics* as seed journal. Citations between journals are normalized
(cosine>0.2) to enhance visualization. Size of node represents number of citations (normalized); x-axis corrects
for self-citations. Thickness of lined/edges represents tie strength.